\documentclass[aps,prd, longbibliography, twocolumn, nofootinbib, floatfix]{revtex4-1}
\usepackage[hyperindex,breaklinks]{hyperref}
\usepackage[anythingbreaks]{breakurl}
\usepackage{amsmath}
\usepackage{graphicx}
\usepackage{dcolumn}
\usepackage{aas_macros}
\usepackage{bm}
\usepackage{epsfig}
\usepackage{amssymb,latexsym,mathrsfs}
\usepackage{graphicx}
\usepackage{color}
\usepackage{verbatim}

\hypersetup{
    colorlinks=true,
    linkcolor=red,
    citecolor=blue,
    urlcolor=magenta,
} 

\newcommand{\be}{\begin{equation}}
\newcommand{\ee}{\end{equation}}
\newcommand{\bea}{\begin{eqnarray}}
\newcommand{\eea}{\end{eqnarray}}

\newcommand{\om}{\Omega_m} 
 
\newcommand{\gam}{\gamma}

\newcommand{\lcdm}{$\Lambda$CDM} 
\newcommand{\fs}{f\sigma_8} 

\begin{document}

\title{Complementarity of Peculiar Velocity Surveys and Redshift Space Distortions for Testing Gravity} 

\author{Alex G. Kim${}^{1}$, Eric V.\ Linder${}^{1,2,3}$} 
\affiliation{
${}^1$Lawrence Berkeley National Laboratory, Berkeley, CA 94720, USA\\ 
${}^2$Berkeley Center for Cosmological Physics, 
University of California, Berkeley, CA 94720, USA\\ 
${}^3$Energetic Cosmos Laboratory, Nazarbayev University, 
Nur-Sultan, Kazakhstan 010000 
}

\date{\today}

\begin{abstract}
Peculiar-velocity surveys of the low-redshift universe have significant 
leverage to constrain the growth rate of cosmic structure and test 
gravity. 
Wide-field imaging surveys combined
with multi-object spectrographs (e.g.\ ZTF2, LSST, DESI, 4MOST)
can use Type Ia supernovae as informative 
tracers of the velocity field, reaching few percent constraints on 
the growth rate $f\sigma_8$ at $z\lesssim0.2$ where density tracers cannot do better than 
$\sim10\%$. Combining the high-redshift DESI survey mapping 
redshift space distortions with a low-redshift 
supernova peculiar velocity survey using LSST and DESI can determine the 
gravitational growth index to  $\sigma(\gamma)\approx0.02$, testing general relativity. 
We study the characteristics needed for the peculiar velocity survey, 
and how its complementarity with clustering surveys improves when going 
from a \lcdm\ model assumption to a $w_0$--$w_a$ cosmology. 
\end{abstract} 

\maketitle

\section{Introduction}

The spatial distribution of large scale structure encodes abundant 
information on the cosmological model. The inhomogeneous clustering 
is matched by motions -- peculiar velocities -- with respect to the 
cosmic expansion, and this also contains important information.
These 
velocities appear, for example, in redshift space distortions (RSD) whose 
measurement is a major focus of galaxy surveys such as the Dark Energy 
Spectroscopic Instrument (DESI \cite{2016arXiv161100036D}). However, RSD maps the 
velocity field in a statistical sense in that velocities 
are not determined per object
but are inferred through the clustering of
objects. 

One can also seek to measure peculiar velocities directly from individual 
objects, and then carry out statistical analysis of the velocity field. 
This, however, requires accurate separation of the cosmic expansion 
redshift from the measured redshift, which can be accomplished through 
measured distances and a tight distance-redshift relation. This is generally 
only practical at low redshifts where the redshift from velocities is not 
negligible compared to the cosmic redshift. Such peculiar velocity (PV) surveys 
have employed distance measurements by the fundamental plane relation 
for elliptical galaxies (e.g.\ 6dFGS \cite{2017MNRAS.471..839A} and TAIPAN \cite{2017PASA...34...47D}), 
and the Tully-Fisher relation for spiral galaxies (e.g.\ WALLABY+WNSHS 
\cite{2008ExA....22..151J}); recently Type Ia supernova (SN) standardized candles have 
been considered \cite{PhysRevLett.99.081301,2008MNRAS.389L..47A,2014MNRAS.444.3926J,2015JCAP...12..033H, 2017JCAP...05..015H, 2019BAAS...51c.140K}. 

In the next decade, the Zwicky Transient
Facility \citep[ZTF,][]{2019PASP..131a8002B}, its proposed successor (ZTF-II), and 
the Large Synoptic Survey Telescope \citep[LSST,][]{2017ApJ...847..128H} will find 
$\sim100,000$ SNe at $z<0.2$, and DESI and 4MOST \cite{2019Msngr.175...58S} can determine accurate redshifts for 
the host galaxies. With such a number density of distance indicators 
accurate to perhaps 4-5\% each, the velocity field can be mapped with good 
signal to noise and the cosmic growth rate $\fs$ measured to the equivalent of $\sim2\%$ in 
each of two redshift bins at $z=[0,0.1]$ and $[0.1,0.2]$ 
(although there is no need to bin). 

This is interesting for cosmology, but especially for testing gravity. 
That is because the velocity field is an 
especially robust way to test gravity: density and velocity are simply 
related by the continuity equation (due to mass conservation), and 
velocity is proportional to acceleration (in linear perturbation theory) 
by Euler's equation for any central force law. Thus these should hold in 
a wide class of gravity theories. Acceleration is related to density 
perturbations through Poisson's equation, and causes the peculiar 
velocities. Thus peculiar velocities can test the strength of gravity. 

In this article we will not assume a particular model of gravity but 
rather constrain deviations from general relativity in terms of the 
gravitational growth index $\gamma$ \cite{2005PhRvD..72d3529L}. The growth index is 
an accurate way of expressing deviations in the growth rate for a wide 
range of gravity theories, as long as they are scale independent on the 
scales of interest (linear theory) and do not affect the early universe 
initial conditions \cite{2007APh....28..481L}. 

Another robust feature of velocity measurements is that the probes
(e.g.\ galaxies) are merely test particles
that lie in and map the velocity field sourced
by all matter, including dark matter.  This in contrast to overdensity measurements, in which the connection between
the baryonic probes (e.g.\ galaxy light) and dark matter is less direct. 

In Sec.~\ref{sec:method} we briefly summarize the survey characteristics 
and method of using the peculiar velocity power spectrum (and 
cross-correlation with the matter density power spectrum) to constrain 
cosmological parameters. We present the results on $\gamma$ and other 
parameters in Sec.~\ref{sec:results}, especially the complementarity with 
higher redshift RSD surveys. We conclude in Sec.~\ref{sec:concl}.

\section{Surveys and Method} \label{sec:method} 

The properties of peculiar velocity
surveys that most strongly determine their
sensitivity to the growth of structure are:
\begin{itemize}
    \item The survey volume: We
    assume a volume of the shape of a shell
    defined by its
    minimum ($r_{\rm min}$) and maximum
    ($r_{\rm max}$) radial distances, and its solid angle 
    ($\Omega$).
    At small galaxy separations,
    observed redshifts can be dominated by peculiar rather than cosmological redshifts, a regime that is not well-described by linear theory: we thus adopt an $r_\text{min}$ that corresponds to $z_\text{min}=0.01$.
    While
    source detections are generally based on  source magnitude, for convenience we set $z_\text{max}=0.2$ up to which complete SN~Ia telescope follow-up is reasonable and beyond which large velocity uncertainties limit the precision in measuring the growth of structure.
    We consider that to cover half of the
    extra-galactic sky, $\Omega = 2\pi$, using both northern 
    and southern hemisphere resources. 
    \item  Number density of the probes $n$:  While different classes of object may be used as density and velocity probes, in this article we consider one class used for both.  For transient probes such as SNe~Ia, the density is directly related to the survey duration.  We consider a 10-year survey with 0.65 efficiency and
    a SN-frame rate of $2.69 \times 10^{-5}(h/0.70)^3\, \text{Mpc}^{-3} \text{yr}^{-1}$ \cite{2010ApJ...713.1026D}.
    \item $\sigma_M$: Intrinsic magnitude dispersion of the probe:   A fixed magnitude dispersion transforms into a distance-dependent velocity dispersion $\sigma$ through $\sigma_M^2 =  \left(\frac{5}{\ln{10}}\right)^2 \left(1-\frac{1}{aH\chi}\right)^2 \sigma^2$, 
    where $\chi$ is the comoving distance.
    We consider a value of $\sigma_M=0.08$~mag \cite{2012MNRAS.425.1007B, 2014ApJ...789...32B, 2015ApJ...815...58F,2018A&A...615A..45S}
    that could be achieved using data beyond optical
    photometry.
    \item Range of $k$ included in the analysis: $k_{\text{min}}$  and $k_{\text{max}}$ are set respectively according to the linear extent of the survey volume and the smallest scales that are confidently modelled. We use $k_{\rm min}=(\pi/r_{\text{max}})\,h$/Mpc 
    and $k_{\rm max}=0.1\,h$/Mpc. 
\end{itemize}

A cosmological model predicts the statistical properties of the spatial distribution and velocities of the probe through
the density-density ($P_{\delta \delta}$), velocity-velocity ($P_{vv}$), and the
cross density-velocity ($P_{\delta v}$) power spectra.
In linear perturbation theory, the dark-matter overdensity
field can be decomposed into independent
temporal and spatial components,
where the temporal component $D(t)$ is known as the growth function.
To first order, 
$P_{vv}\propto (fD\mu)^2$, the SN~Ia host-galaxy count overdensity
power spectrum $P_{\delta\delta}\propto (bD + fD\mu^2)^2$, and the galaxy-velocity cross-correlation $P_{v\delta}
\propto  (bD + fD\mu^2)fD\mu$ 
\cite{2009JCAP...10..004S}, where 
$f=d\ln D/d\ln a$, $b$ is the bias between the
SN hosts and dark matter, and the angle between
the lines of sight toward the
two galaxies is given by
$\mu=\cos(\hat{\mathbf{r}}_1 \cdot \hat{\mathbf{r}}_2)$.
Note that the commonly used
mass fluctuation amplitude
$\sigma_8$ is proportional to
$D$.

We consider two approaches to extracting cosmological information 
from the power
spectra.  For the first approach, $fD$ is
taken to have constant, independent values in
a set of redshift bins, and $bD$ is taken to have a constant
value over all redshift bins  (stable clustering). 
The parameter set is then $\lambda \in \{fD_1,\ldots,fD_{n_b}, bD\}$ for $n_b$ redshift bins.
This  model is commonly
used
to project and report the results of peculiar velocity surveys \cite{2017MNRAS.464.2517H}.

The second approach puts the focus on gravity. 
References~\cite{2005PhRvD..72d3529L,2007APh....28..481L} found that
$f=\om(a)^\gamma$ provides a highly accurate ($\lesssim0.3\%$) 
description of the growth of structure, where a wide range 
of gravity models can be described by single
values of the growth index $\gamma$.
The mass density in units of the critical density, $\om(a)$,  itself
depends on cosmological parameters describing the 
background expansion; for this article we consider
a flat cosmology with the standard $w(a)=w_0 + w_a(1-a)$ dark energy equation
of state.  The bias $b$ is taken to be constant in the narrow $0<z<0.2$
redshift range.
The density and velocity covariances then depend on the parameter set $\lambda \in \{\gamma, \om, w_0, w_a, b\}$,
where $\om$ (without an argument) is the mass density today.

We project parameter uncertainties\footnote{The code is available
at \url{https://github.com/LSSTDESC/SNPeculiarVelocity}.} using the
Fisher information matrix
\begin{multline}
F_{ij} 
 = \frac{\Omega}{8\pi^2} \int_{r_{\rm min}}^{r_{\rm max}} dr  \int_{k_{\rm min}}^{k_{\rm max}} dk \int_{-1}^{1} d\mu\,r^2
  k^2 \\ \text{Tr}\left[ C^{-1} \frac{\partial C}{\partial \lambda_i} C^{-1} 
\frac{\partial C}{\partial \lambda_j} \right]  
\label{fisher:eqn}
\end{multline}
where
\begin{equation}
C(k,\mu,a)  =
  \begin{bmatrix}
   P_{\delta \delta}(k,\mu,a) + \frac{1}{n} &
   P_{v\delta}(k,\mu,a)  \\
   P_{v\delta}(k,\mu,a)  &
  P_{vv}(k,\mu,a) + \frac{\sigma^2}{n}
   \end{bmatrix}.
\label{cov:eq}
\end{equation} 
The fiducial cosmology is a $\Lambda$CDM model ($w_0=-1$,
$w_a=0$), with $\om=0.3$, and a power
spectrum calculated by
CAMB \cite{Lewis:2002ah}
using its default configuration.

\section{Results Testing Gravity} \label{sec:results} 

From the peculiar velocity survey we have measurements of $\fs$ over $0.01<z<0.2$. We can compare or combine these with measurements 
of redshift space distortions 
from DESI. The precisions that DESI 
is expected to deliver are listed in Tables~2.3 and 2.5 of \cite{2016arXiv161100036D}. 
Since our main goal here is testing gravity, we only use its $\fs$ 
measurements (with $k_{\rm max}=0.1\,h$/Mpc), not the baryon acoustic 
oscillation measurements. We also 
emphasize that DESI provides far more cosmological leverage than the 
growth index $\gamma$ constraints we focus on here; in particular, measuring the cosmic growth history 
over a wide range of redshift as DESI does  
is highly fundamental and insightful. 

To motivate why low-redshift surveys can be competitive with the huge 
volumes of high-redshift surveys for the particular goal of measuring 
$\gam$, note that as one approaches the matter dominated era the 
growth rate $f$ approaches unity. Since $\gam$ is 
defined through $f=\om(a)^\gam$, then $\gamma$ is relatively poorly 
determined as $\om(a)\to1$ at higher redshifts. Since $\fs\propto fD$ and $D/a\to1$ 
(appropriately normalized) at higher redshifts, the insensitivity to 
$\gam$ is compounded. However, at low redshift, both $f$ and $D/a$ 
deviate from unity and $\gam$ has increased influence, making low 
redshift measurements of $\fs$ a good avenue for testing whether $\gam$ 
is consistent with its general relativity value of 0.55. 

\begin{figure}[htbp!] 
\centering
\includegraphics[width=\columnwidth]{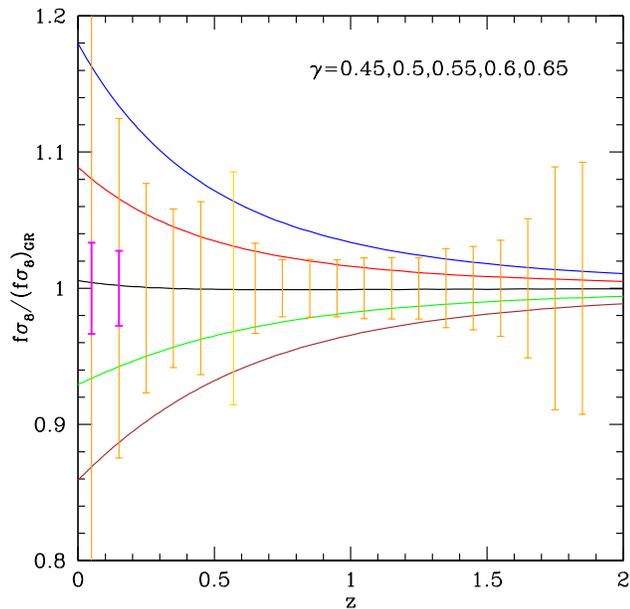} 
\caption{
The ratio of the growth rate $\fs$
to the general relativity (GR) 
behavior is plotted for five values of the gravitational growth 
index $\gamma$, increasing from top to bottom. Note that $\gam=0.55$ 
is a highly accurate 
approximation to GR. DESI RSD (medium orange), current BOSS RSD (lighter gold), and peculiar velocity survey (dark magenta)  
uncertainties in 0.1 redshift bins are overplotted. 
} 
\label{fig:fsband}
\end{figure}

Figure~\ref{fig:fsband} illustrates this, plotting $\fs(z)$ for 
various values of $\gam$, relative to the general relativity 
behavior for the same background cosmology. The 
curves flare out at low redshift, pointing to 
an opportunity there if the measurement precision can be made 
reasonable. Overplotted are the expected measurement uncertainties 
on $\fs$ in redshift bins of width 0.1 
from the DESI RSD in the main survey (above $z>0.6$), 
from the DESI RSD in the Bright Galaxy Survey (BGS; below $z<0.5$), 
and our baseline peculiar velocity survey, using the first 
approach of Sec.~\ref{sec:method}.

\begin{figure}[htbp!] 
\centering
\includegraphics[width=\columnwidth]{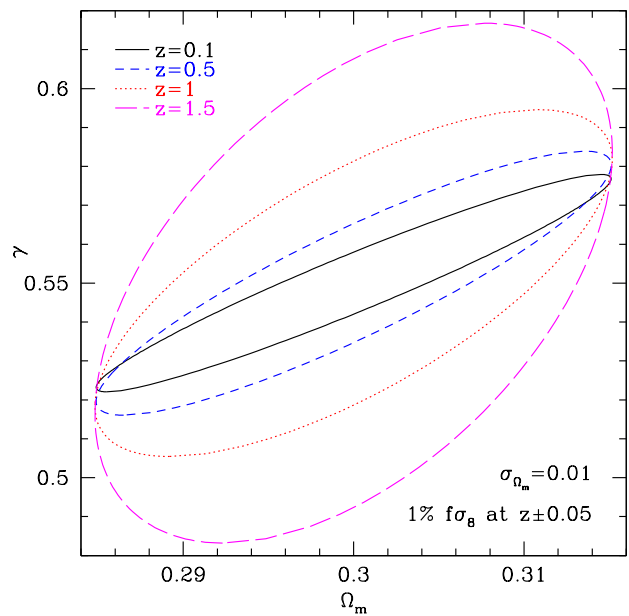} 
\caption{
Low-redshift measurements of the growth rate $\fs$ give improved constraints 
in the $\om$--$\gam$ plane. Relative to higher redshift, $z\approx0.1$ 
measurements give tighter confidence contours and ones oriented more 
narrowly in the $\gamma$ axis. This plot is for illustrative 1\% 
measurements of $\fs$ at $z\pm0.05$ in a \lcdm\ universe, combined with 
a Gaussian prior of 0.01 on $\om$, at four different redshifts $z$. 
} 
\label{fig:gamflowerp}
\end{figure}

While the $\gam$ curves for $\fs(z)$ give a useful indication of 
the strength of deviations from general relativity for a given 
$\gam$, the observational constraints on $\gam$ will also depend on 
covariance with other cosmological parameters such as the matter 
density $\om$. The same advantages of low redshift measurements 
still hold, demonstrated in 
Fig.~\ref{fig:gamflowerp} 
with the joint confidence contours 
on $\gamma$ and $\Omega_m$. This gives a feel for the sensitivity 
of measurements near particular redshifts, by adopting simply 
a localized pair of measurements at $z\pm0.05$. 
(A Gaussian prior of 0.01 on $\om$ is included for convenience in drawing  reasonable 
confidence contours from only two measurements; we have 
checked that this does not substantially alter the degeneracy direction.) 

Keeping the measurement 
precision fixed (at 1\% for purely illustrative purposes), high-redshift measurements of $\fs$ have less constraining 
power on $\gam$ than low-redshift ones. 
Furthermore, the degeneracy 
direction of the contours slowly rotates with the measurement redshift, 
becoming more favorable with respect to determining $\gam$ at low 
redshift. 
The general 
degeneracy direction is simple to understand: increasing $\om$ increases 
growth, while increasing $\gam$ (essentially decreasing the strength of 
gravity; see \cite{2007APh....28..481L}) decreases growth. Thus doing both compensates 
for each, extending the contour along the positive slope diagonal. At 
high redshift, since the measurement is less sensitive to $\gam$, a given 
change in $\om$ requires a larger change in $\gam$, steepening the slope 
and rotating the contour counterclockwise; for low redshift, the opposite 
occurs and the clockwise-rotated contour gives tighter constraints on $\gam$.

Having established that low-redshift peculiar velocity surveys have 
the potential to help test gravity, we now carry out a full Fisher 
information matrix analysis of the cosmological and gravity parameter constraints 
(using the second approach of Sec.~\ref{sec:method} and so 
without any need to bin 
measurements of $f\sigma_8$) 
enabled by the baseline PV survey of Sec.~\ref{sec:method}, and also that 
combined with DESI RSD measurements. We consider two cosmological models: 
\lcdm\ characterized by the matter density $\om$, and dynamical dark energy 
characterized by $\om$ and the dark energy parameters $w_0$ and $w_a$ 
describing its present equation of state and its time variation. In 
addition there is the gravitational growth index $\gamma$ (and 
source bias parameter, which is always marginalized over). To roughly 
represent other cosmological data besides PV and RSD in the 
DESI-LSST era, we impose a 
Gaussian prior on $\om$ of width 0.01. 

We find that in the \lcdm\ model, PV alone determines $\gam$ to 0.019 
and RSD alone to 0.026. That is, within this limited focus on testing 
gravity in the form of $\gamma$ (which indeed provides a subpercent 
accurate fit to the effect of many modified-gravity models on cosmic 
growth in the linear regime), low-redshift PV can match DESI RSD. 
Combining the two yields an improvement to $\sigma(\gam)=0.018$. 
Recall that the distance in $\gam$ between general relativity and $f(R)$ gravity 
or braneworld gravity is $\pm0.13$ respectively. 

\begin{figure}[htbp!] 
\centering
\includegraphics[width=\columnwidth]{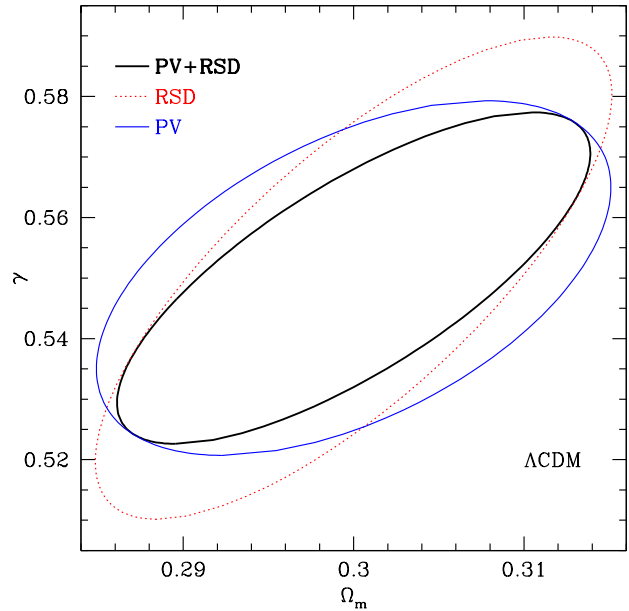} 
\caption{
Peculiar velocity (PV) measurements from our fiducial survey at $z=0$--$0.2$ 
give comparable constraints (contours at the $1\sigma$ joint confidence 
level) in the $\om$--$\gam$ plane to DESI redshift 
space distortion (RSD) measurements over $z=0$--1.8. (Of course DESI 
gives crucial information on the whole growth history, not merely its 
compression into $\om$--$\gam$.) The combination of the two surveys gives 
further improvement. Here \lcdm\ is assumed. 
} 
\label{fig:omgell}
\end{figure}

Figure~\ref{fig:omgell} shows the $1\sigma$ joint confidence contours 
in the $\om$--$\gam$ plane for these cases. 
We see that the two dimensional joint confidence contours 
of PV and RSD are complementary, and the ``figure of 
merit'' (inverse area in terms of the inverse square root of the determinant of 
the $\om$--$\gam$ covariance matrix) of the 
combined confidence region improves by 1.5/1.6 times relative 
to the individual probes, respectively.

However, when we allow more cosmological freedom as in the $w_0$--$w_a$ 
model, the situation is different. The constraints become 
$\sigma(\gam)=0.15$ for DESI RSD but 0.028 for PV+RSD, i.e.\ peculiar 
velocities play a crucial role in testing gravity when there is also 
freedom in the nature of dark energy. Essentially, the PV survey  together with the RSD survey allows 
simultaneous fitting of $\om$, $w_0$, $w_a$, and $\gam$ with reasonable 
constraints. 

\begin{figure}[htbp!] 
\centering
\includegraphics[width=\columnwidth]{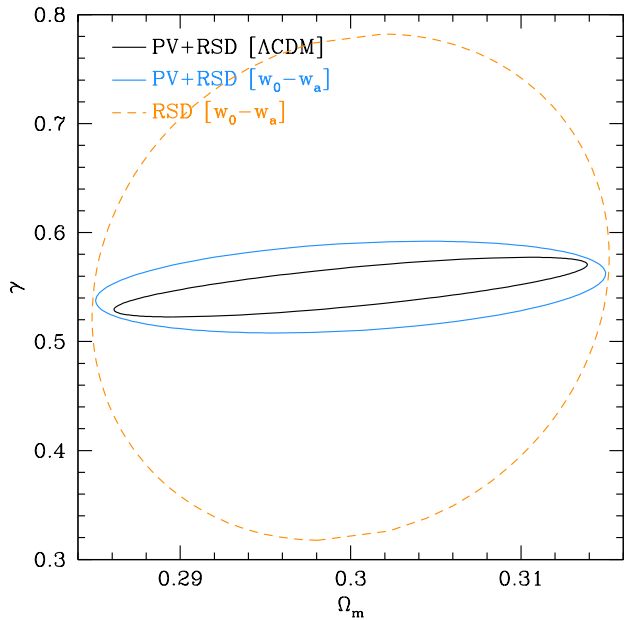} 
\caption{
As Fig.~\ref{fig:omgell}, but with marginalizing over the dark energy 
equation of state parameters $w_0$, $w_a$ rather than fixing to \lcdm. 
Note the complementarity between peculiar velocities and redshift space 
distortions keeps the constraints from weakening too much when also 
fitting for $w_0$ and $w_a$ (solid light blue RSD+PV contour vs the solid black 
contour fixed to \lcdm\ and the same as in Fig.~\ref{fig:omgell}). 
However, the confidence contour using RSD alone (dashed orange) 
blows up considerably in going from a \lcdm\ to $w_0$-$w_a$ model. 
} 
\label{fig:omgw0waell}
\end{figure}

Figure~\ref{fig:omgw0waell} shows that RSD alone has 
difficulty fitting all the parameters together, but adding PV substantially 
immunizes against this issue, still allowing reasonable constraints. Now the figure of merit improvement in the 
$\om$--$\gam$ plane for the 
combined probes is a factor of 2.6/5.8 respectively. 
Alternately, one could say that allowing going beyond a 
\lcdm\ background blows up the $\om$--$\gam$ contour for 
RSD alone by a 
factor 8.9, but increases the area for PV+RSD by only a 
factor 2.4, as shown by the black contour going to the blue one.

If modeling allows accurate use of modes out to 
$k_{\rm max}=0.2\,h$/Mpc instead, then the combination of PV+RSD can 
determine $\gam$ to 0.015 (0.019) in the \lcdm\ 
($w_0$--$w_a$) cosmology case. The figure of merit 
improvement in the 
$\om$--$\gam$ plane with the higher $k_{\rm max}$ is a factor of 2.0 (1.5). 
All the results are summarized in Table~\ref{tab:gammas}. 

\begin{table}[htbp!]
\begin{tabular}{|l|c|l|r|l|r|}
 \hline
& & \multicolumn{2}{c|}{$k_\text{max} =0.1$} & \multicolumn{2}{c|}{$k_\text{max} =0.2$} \\ 
\cline{3-6}
Model\    & Data  & \ $\sigma(\gamma)$ & \ FOM \ & \ $\sigma(\gamma)$ & \ FOM \ \\ 
\hline 
\lcdm\ &  PV & \ 0.0193 \ & 6064 \ & \ 0.0156 & 8288 \ \\ 
\lcdm\ & RSD & \ 0.0262 & 5891 \  & \ 0.0214 & 12267 \ \\ 
\lcdm\ & PV+RSD & \ 0.0180 & 9208 \  & \ 0.0153 & 17975 \ \\ 
$w_0$-$w_a$\ & PV & \ 0.0691 & 1466 \ & \ 0.0491 & 2088 \ \\ 
$w_0$-$w_a$\ & RSD & \ 0.153 & 659 \  & \ 0.0707 & 1467 \  \\
$w_0$-$w_a$\ & PV+RSD & \ 0.0277  & 3828 \  & \ 0.0195 & 5852 \ \\ 
\hline
 \end{tabular}  
\caption{
Summary of constraints on the gravitational growth index $\gamma$ and the 
figure of merit (FOM: $1/\sqrt{\det COV[\om,\gamma]}$) for various 
cosmology models and data sets. PV is the low redshift peculiar velocity 
survey, RSD is the DESI redshift space distortions. We show both the 
baseline case of 
using modes out to $k_{\rm max}=0.1\,h/$Mpc and the optimistic case with  
$0.2\,h$/Mpc. All cases include a prior of $\sigma(\om)=0.01$. 
} 
\label{tab:gammas} 
\end{table} 

\section{Conclusions} \label{sec:concl} 

Peculiar velocities provide a direct measurement
of the growth of structure, and hence serve as a
powerful probe of the
gravitational forces responsible for the clustering and
motion within the expanding Universe.
At low redshift, peculiar velocities can be measured
from individual galaxies whose distances are accurately
measured, e.g.\ through Type Ia supernovae.  At high redshift, the imprint of peculiar velocities creates a specific anisotropy in the correlation
functions of the redshift-space coordinates
of ensembles of galaxies.

Each approach can provide constraints on the growth
index $\gamma$ that can distinguish classes of gravity.
A distinguishing feature of direct peculiar velocities
is that precise measurements are confined to
low redshift, but here the sensitivity to $\gamma$ ``flares'' in  
enhancement, while higher redshift RSD 
measurements are more sensitive to $\om$ than $\gamma$ (since 
as $\om(a)\to1$ then $\gamma$ is poorly determined). The two 
methods therefore have great complementarity in testing gravity. 
This advantage strengthens further when
$\om(a)$ becomes more flexible in models that
go beyond $\Lambda$CDM. 

The upcoming generation of redshift and transient object 
surveys will give exciting data sets that can be used in 
these ways. Cadenced
multi-band imaging surveys
such as ZTF-2 and LSST can
provide near-full sky, complete 
low-redshift
SN discoveries and light curves used to obtain host distances.
Wide-field multi-object
spectrographs such as DESI
and 4MOST can follow-up these
discoveries to get multiplexed transient
classifications and 
redshifts; the use of these 
instruments for peculiar velocities
is being considered by
the DESI Collaboration and in
the 4MOST Hemisphere Survey
proposal.  Other 2-4m class
facilities can
provide supplemental near infrared and spectroscopic data to  improve the data set
\cite{2004SPIE.5249..146L,2019eeu..confE..42R}.  A coordinated
effort is required to organize the
diverse range of facilities
that comprise a complete
peculiar-velocity survey. 

We projected cosmology and gravity constraints with two 
approaches, using $\fs$ and going directly to the gravitational 
growth index $\gamma$. The Fisher code for dealing with peculiar 
velocity surveys is publicly available. The results highlight 
the significant complementarity between peculiar velocity 
and redshift space distortion surveys, especially as the 
dark energy properties are fit simultaneously. Testing gravity 
with $\sigma(\gamma)\approx0.02$ appears achievable. Further 
improvements are possible if our understanding of perturbation 
theory allows use of smaller scale modes, with figure of merit 
increases of 1.5--2. Any sign of deviation of the value of $\gam$ 
from general relativity would then motivate more detailed analysis 
with more sophisticated tests of how gravity behaves on cosmic 
scales.

\section*{Acknowledgments}

We thank Arman Shafieloo and KASI for hospitality during part of this work and 
Yong-Seon Song for helpful conversations. 
This work is supported in part by 
the U.S.\ Department of Energy, Office of Science, Office of High Energy 
Physics, under Award DE-SC-0007867 and contract no.\ DE-AC02-05CH11231, 
and by the Energetic Cosmos Laboratory. 
\bibliography{alex}

\end{document}